\documentclass[12pt]{article}
\usepackage{amssymb}

\usepackage{amsmath}
\usepackage{graphicx, epsfig}

\begin{document}

\title{Tomographic analysis of reflectometry data II: the phase derivative}
\author{{\small Fran\c{c}oise Briolle}\thanks{%
Corresponding author, francoise.briolle@univmed.fr} \thanks{%
Centre de Physique Th\'{e}orique, CNRS Luminy, case 907, F-13288 Marseille
Cedex 9, France}, {\small Ricardo Lima}\footnotemark[2] , {\small and Rui
Vilela Mendes}\thanks{%
IPFN - EURATOM/IST Association, Instituto Superior T\'{e}cnico, Av. Rovisco
Pais 1, 1049-001 Lisboa, Portugal} \thanks{%
CMAF, Complexo Interdisciplinar, Universidade de Lisboa, Av. Gama Pinto, 2 -
1649-003 Lisboa, Portugal, e-mail: vilela@cii.fc.ul.pt}}
\maketitle

\begin{abstract}
A tomographic technique has been used in the past to decompose complex
signals in its components. The technique is based on spectral decomposition
and projection on the eigenvectors of a family of unitary operators. Here
this technique is also shown to be appropriate to obtain the instantaneous
phase derivative of the signal components. The method is illustrated on
simulated data and on data obtained from plasma reflectometry experiments in
the Tore Supra.
\end{abstract}



\section{Introduction: Plasma density from reflectometry and its
multicomponent nature}

Density measurements play an important role in the study and operation of
magnetically confined plasmas. Microwave reflectometry is a radar-like
technique which infers the plasma density from the reflection on the
(cutoff) layers where the refractive index vanishes \cite{Simonet}. For
example, for propagation perpendicular to the magnetic field with the
electric field of the wave parallel to the magnetic field in the plasma
(O-mode), the refractive index is 
\begin{equation}
\mu =\sqrt{1-\frac{\omega _{p}^{2}}{\omega ^{2}}}=\sqrt{1-\frac{n_{e}e^{2}}{%
\varepsilon _{0}m_{e}\left( 2\pi f\right) ^{2}}}  \label{1.1}
\end{equation}
where $n_{e}$ is the electron density, $\omega _{p}$ $=\left( \frac{4\pi
n_{e}e^{2}}{m_{e}}\right) ^{\frac{1}{2}}$ the plasma frequency, $e$ and $%
m_{e}$ the electronic charge and mass, $\varepsilon _{0}$ the permittivity
of the vacuum and $f=\frac{\omega }{2\pi }$ the frequency of the probing
wave. When the plasma frequency equals the probing frequency the index of
refraction vanishes, the wave is reflected and the density $n_{c}$ of the
cutoff layer may be derived from 
\begin{equation}
n_{c}=\frac{\varepsilon _{0}m_{e}\left( 2\pi f\right) ^{2}}{e^{2}}
\label{1.2}
\end{equation}
Mixing the reflected wave $E_{R}\left( t\right) $ with the (reference)
incident wave $E_{0}\left( t\right) $, the mixer output is 
\begin{equation*}
\frac{1}{2}\left( E_{0}^{2}\left( t\right) +E_{R}^{2}\left( t\right) \right)
+E_{0}\left( t\right) E_{R}\left( t\right) \cos \phi \left( t\right) 
\end{equation*}
In the interference term $E_{0}\left( t\right) E_{R}\left( t\right) \cos
\phi \left( t\right) $,  $E_{0}\left( t\right) E_{R}\left(
t\right) $ depends on many factors, microwave generator power, plasma
scattering properties, turbulence, etc., therefore it is $\phi \left(
t\right) $ that contains the most reliable information about the plasma
density.

The location $x_{c}\left( \omega _{p}\right) $ of the reflecting layer for
the frequency $\omega _{p}$ is related to the group delay 
\begin{equation}
\tau _{g}=\frac{d\phi \left( \omega \right) }{d\omega }=\frac{1}{2\pi }\frac{%
d\phi }{df}  \label{1.3}
\end{equation}
by (O-mode) 
\begin{equation}
x_{c}\left( \omega _{p}\right) =x_{0}+\frac{c}{\pi }\int_{0}^{\omega
_{p}}d\omega \frac{1}{\left( \omega _{p}^{2}-\omega ^{2}\right) ^{\frac{1}{2}%
}}\frac{d\phi \left( \omega \right) }{d\omega }  \label{1.4}
\end{equation}
For a linear frequency sweep of the incident wave 
\begin{equation}
f\left( t\right) =f_{0}+\gamma t  \label{1.5}
\end{equation}
one obtains 
\begin{equation}
\frac{d\phi }{df}=\frac{1}{\gamma }\left. \frac{d\phi }{dt}\right\vert _{f}
\label{1.6}
\end{equation}
Therefore, measurement of the plasma density hinges on an accurate
determination of the \textquotedblleft instantaneous
frequency\textquotedblright\ $\frac{d\phi }{dt}$. Accuracy in the
measurement of this quantity is quite critical because, the location of the
reflecting layer being obtained from the integral in (\ref{1.4}), errors
tend to accumulate.

Several methods have been devised to obtain the group delay $\tau _{g}$ from
the reflectometry data (for a review see \cite{Manso1}). Among them,
time-frequency analysis \cite{Manso3} has been, so far, the most promising
technique. The Wigner-Ville (WV) distribution \cite{Manso2} \cite{Bizarro}
although providing a complete description of the signal in the
time-frequency plane, raises difficult interpretation problems due to the
presence of many interference terms that impair the readability of the
distribution. This occurs because the VW (quasi-)distribution is not a
probability distribution, has complex amplitudes and may have large
amplitude values in frequency regions which are not contained in the signal
spectrum. For this reason the time-frequency method that has been preferred
is the spectrogram \cite{Manso3} \cite{Clairet} \cite{Clairet2} \cite{Manso4}%
, that is, the squared modulus of the short-time Fourier transform 
\begin{equation*}
SP\left( t,f\right) =\left\vert \int_{-\infty }^{\infty }x\left( u\right)
h\left( u-t\right) e^{-i2\pi fu}du\right\vert ^{2} 
\end{equation*}
$h\left( u\right) $ being a peaked short-time window.

The spectrogram does not really provide the instantaneous frequency, because
that notion is not well defined anyway. All it gives is the product of the
spectra of $x\left( t\right) $ and $h\left( t\right) $. The way the
spectrogram is used to infer the local rate of phase variation $\frac{d\phi 
}{dt}$ is to identify this quantity with the maximum or the with the first
moment of the spectrogram. An additional problem comes about because
unwanted phase contributions due to plasma turbulence may have a higher
amplitude than the contributions due to the profile. Correction techniques
have been developed to compensate for this errors, based for example on
Floyd's best path algorithm. The choice of the window function is also an
important issue and, in particular, an adaptive spectrogram technique has
been developed to maximize the time-frequency concentration \cite{Manso1}.

In addition to the delicate nature of the extraction of the phase derivative
from an interference signal, another important question is the
multicomponent structure of this signal. The signal that is actually
received contains, in addition to the reflection on the plasma, reflections
on the porthole and multi reflections of the waves on the wall of the vacuum
vessel. Separation of these latter components from the plasma reflections is
an essential step to obtain reliable density results. Separation by
frequency filtering is not appropriate because there is considerable
frequency overlap between the components. In a previous paper\cite{Briolle1} we
have developed a method to separate the signal components based on a
tomographic representation\cite{PLA} \cite{JPA}, which gives a positive
density $M_{f}(x,\theta )$\ of the signal along all possible $\theta $%
-directions in the time-frequency plane.

\ The tomogram representation $M_{f}(x,\theta )$ gives, for $\theta =0$ the
time representation of the signal, $f(t)$, and for $\theta =\frac{\pi }{2}$
the frequency representation, $\overset{\sim }{f}(\nu )$. Frequency
filtering corresponds to component separation of the signal at $\theta =%
\frac{\pi }{2}$ and, from the many examples that were studied, one concludes
that, in general, this is not the most convenient direction to isolate the
signal components. For example, for the reflectometry signals that were
studied, we have more information if the separation of the components is
performed at $\theta =\frac{3\pi }{10}$ than at $\theta =\frac{\pi }{2}$.

In the next section we first make a brief review of the tomographic method
for component separation and then, using the same mathematical framework,
show how one can obtain the phase derivative from the isolated components.

Finally, in the last two sections, the methods are applied both to simulated
data and to actual reflectometry data collected in the Tore Supra.

\section{Tomograms, components and the phase derivative}

In \cite{Briolle1} we described in much detail the use of tomograms for the
component factorization of complex signals. Here we just recall some basic
facts for the reader's convenience.

We define a (time-frequency\footnote{%
As explained in \cite{Briolle1}, other non-commuting operator pairs may be
chosen}) tomogram as a family of probability distributions, $M_{f}(x,\theta
) $ associated to any signal $f(t)$, $t\in \lbrack 0,T]$ by

\begin{equation}
M_{s}(x,\theta )=\left\vert \int \ f(t)\Psi _{x}^{\theta
,T}(t)\,dt\right\vert ^{2}=\left\vert <f,\Psi _{x}^{\theta ,T}>\right\vert
^{2}  \label{2.1}
\end{equation}
with 
\begin{equation}
\Psi _{x}^{\theta ,T}(t)=\frac{1}{\sqrt{T}}\exp \left( \frac{-i\cos \theta }{%
2\sin \theta }\,t^{2}+\frac{ix}{\sin \theta }\,t\right)  \label{2.2}
\end{equation}
Notice that $\left\vert \Psi _{x}^{\theta ,T}><\Psi _{x}^{\theta
,T}\right\vert $ are spectral projections of an unitary operator $U(\theta )$
and therefore (\ref{2.1}) performs a spectral decomposition of the signal.

First we select a subset of numbers $\left\{ {x_{n}}\right\} $ in such a way
that the corresponding family $\left\{ \Psi _{x_{n}}^{\theta ,T}(t)\right\}
_{n}$ is orthogonal and normalized:

\begin{equation}
<\Psi _{x_{m}}^{\theta ,T},\Psi _{x_{n}}^{\theta ,T}>=\delta _{m,n}
\label{2.3}
\end{equation}

A glance at the shape of the functions (\ref{2.2}) shows that, for fixed $%
\theta $, the oscillation length at a given $t$ decreases when $\left\vert
x\right\vert $ increase. As a result, the projection of the signal on the $%
\left\{ \Psi _{x_{n}}^{\theta ,T}(t)\right\} $ basis locally explores
different scales. On the other hand the local time scale is larger when $%
\theta $ also becomes larger, in agreement with the uncertainty principle
for a non-commuting pair of operators.

We then consider the projections of a signal $f(t)$

\begin{equation}
c_{x_{n}}^{\theta }(f)=<f,\Psi _{x_{n}}^{\theta ,T}>  \label{2.6}
\end{equation}
and use the coefficients $c_{x_{n}}^{\theta }(f)$ for our signal processing
purposes.

In particular, a multi-component analysis of the signal\cite{Briolle1} is
done by selecting subsets $\mathcal{F}_{k}$ of the $\left\{ x_{n}\right\} $
and reconstructing ($k$-component) partial signals by restricting the sum to

\begin{equation}
f_{k}(t)=\sum_{n\in \mathcal{F}_{k}}c_{x_{n}}^{\theta }(f)\Psi
_{x_{n}}^{\theta ,T}(t)  \label{2.8}
\end{equation}
for each $k$-component.

In the present work we analyze the phase derivative of a complex signal $%
f(t)=A(t)e^{i\phi (t)}$ and consider the cases where $f(t)$ already
corresponds to one of the components determined as in \cite{Briolle1}. That
is, after a convenient factorization of the signal is performed, the search
for the phase derivative is made on each component.

In the reflectometry technique the experimental signal is already complex
(it consists of one recorded interference term composed of in-phase and $%
90^{o}$ phase shifted signals). Therefore we have no ambiguity in the
definition of the amplitude and phase of $f(t)$. For other type of signals,
where only the real part is available, the construction of a complementary
imaginary part is an usual technique for which there are standard methods
available in the signal analysis literature (see \cite{Boashash} and
references therein).

Given a signal $f(t)=A(t)e^{i\phi (t)}$ the time derivative of the phase may
be obtained from%
\begin{equation}
\frac{\partial }{\partial t}\phi (t)=\textit{Im}\left( \frac{\frac{\partial f}{%
\partial t}}{f\left( t\right) }\right)   \label{2.9a}
\end{equation}%
For our decomposed components one has

\begin{equation}
\frac{\partial }{\partial t}\phi (t)=\textit{Im}\left( \frac{\Upsilon (t)}{%
\tilde{y}(t)}\right)   \label{2.9}
\end{equation}%
with 
\begin{equation}
\Upsilon (t)=\sum_{x_{n}}c_{x_{n}}^{\theta }(f)\frac{\partial }{\partial t}%
\Psi _{x_{n}}^{\theta ,T}(t)  \label{2.10}
\end{equation}%
and 
\begin{equation}
\tilde{y}(t)=\sum_{x_{n}}c_{x_{n}}^{\theta }(f)\Psi _{x_{n}}^{\theta ,T}(t)
\label{2.11}
\end{equation}

Notice that an explicit analytic expression for $\frac{\partial}{\partial t}
\Psi^{\theta,T}_{x_n} (t)$ is known, namely:

\begin{equation}
\frac{\partial }{\partial t}\Psi _{x_{n}}^{\theta ,T}(t)=i\left( \frac{-\cos
\theta }{\sin \theta }t+\frac{x}{\sin \theta }\right) \Psi _{x_{n}}^{\theta
,T}(t)  \label{2.12}
\end{equation}%
and therefore we obtain a direct expression for the phase derivative in
terms of the coefficients $c_{x_{n}}^{\theta }(f)$ without having to use the
values of $f$ for neighboring values of $t$. This provides a more robust
method to estimate the derivative. We call the \textbf{Tomographic Direct
Method (TDM)} the method of the computation of the phase derivative of $f(t)$
using (\ref{2.9}).

Notice that in the calculation of the imaginary part in (\ref{2.9a}) the
value of the amplitude $A\left( t\right) $ plays no role. Therefore we may
use what will be called a \textbf{Tomographic Normalized Method (TNM)}
defined in the same way as \textbf{TDM} but with a normalized signal $\frac{%
f(t)}{\left\vert f(t)\right\vert }$ replacing $f(t)$. For calculations on
the signal carried out with absolute precision the results of \textbf{TDM }%
and\textbf{\ TNM} should coincide. However because of  numerical errors, 
normalization of the signal amplitude, before further processing, might
have some merit mostly in the small amplitude regions. 

There are still two specific issues to be addressed when dealing with the
reconstruction of the phase derivative of $f(t)$. The first is a general
problem in signal analysis, namely denoising. We recall from \cite{Briolle1}
that \textbf{Tomogram-Based Denoising (TBD)} consists in eliminating from (%
\ref{2.11}) the $c_{x_{n}}^{\theta }(f)$ such that

\begin{equation}
\left\vert c_{x_{n}}^{\theta }(f)\right\vert ^{2}\leq \epsilon  \label{2.13}
\end{equation}
for some chosen threshold $\epsilon $.

Another way, often used for denoising, consists in locally smoothing the
signal by computing a \textbf{Local Mean ($LM_{m}$)} $G$ of a function $g(t)$%
, known in the signal processing community as \textbf{moving average FIR
filter} of order $2m+1$ by:

\begin{equation}
G(t_{0})={\sum_{k=-m}^{m}\frac{g(t_{0}-t_{k})}{2m+1}}  \label{2.14}
\end{equation}

The second issue is how to handle the difficult problem of accurate
measurement of the phase, hence also of the phase derivative, when the
signal amplitude is very small. Given a complex signal $f(t)$ we define the 
\textbf{\ truncated Phase Derivative (tPD)} by

\begin{equation}
\frac{\partial^T}{\partial t}\phi(t)=0  \notag  \label{2.15}
\end{equation}

if ${f}(t) < \alpha$, for some convenient threshold $\alpha$,

else

\begin{equation}
\frac{\partial ^{T}}{\partial t}\phi (t)=\textit{Im}\left( \frac{\Upsilon (t)}{%
\tilde{y}(t)}\right)   \label{2.16}
\end{equation}%
Notice that \textbf{tPD} simply sets the value of the phase equal to a
constant when the signal amplitude prevents its accurate estimation.

In the following sections we present some advantages and drawbacks of these
tools by applying them to several simulated and experimental signals.

\section{Examples: Simulated data}

In this section we apply the general method to two types of simulated
signals. The first example shows how the phase derivative of a sinusoidal
signal may be computed with accuracy, even when noise is present. In the second example, we
focus on the phase derivative of signals with non linear phase.

Our data consists of complex functions $y(t)=A(t)e^{i\phi (t)}$ with phase
and phase derivative $\partial _{t}\phi (t)$ unambiguously defined. The
analysis of all the simulated signals is based on tomograms with $\theta =%
\frac{\pi }{5}$ as for the same data in \cite{Briolle1}.

For the simplest case, the signal is:

\begin{equation}
y(t)=\text{exp}(i75t),t\in \left[ 0,20\right] s  \label{3.1.1}
\end{equation}
\textbf{TDM} alone gives an excellent result (mean value of the computed $%
\partial _{t}\phi (t)$ is $74.8rd/s$ and the standard deviation (sdev) $%
0.3rd/s$ to be compared with the Fourier Transform for which the resolution
is equal to $\Delta f=\frac{2\pi }{T}\approx 0.31rd/s$.

If we add a (complex) noise $b(t)$ to (\ref{3.1.1}) with $SNR=10dB$\footnote{%
The SNR is defined by : $SNR(y,b)=10\log _{10}\frac{P_{y}}{P_{b}}$ with $%
P_{y}=\frac{1}{T}\int_{0}^{T}\left\vert y(t)\right\vert ^{2}dt$ and $P_{b}=%
\frac{1}{T}\int_{0}^{T}\left\vert b(t)\right\vert ^{2}dt$.}, \textbf{TDM},
not surprisingly, still shows a good mean result (75.9 rd/s) but has a large
uncertainty (sdev=40 rd/s). The use of \textbf{LM} alone is not sufficient
in this case (sdev=3.5 rd/s for a \textbf{$LM_{15}$}) but \textbf{TBD}
allows a \textbf{TDM} with great accuracy (sdev=0.8 rd/s) that may even be
improved by a ultimate use of a \textbf{LM}, (sdev=0.6 rd/s for \textbf{$%
LM_{5}$}).

It is worthwhile to mention how denoising using the spectral decomposition
of the operator $U(\alpha )$ (\textbf{TBD}) works so efficiently, a result
that is also confirmed in the subsequent examples.

We  proceed to the analysis of a signal which aims to mimic, in a
simplified way, the case of an incident plus a reflected wave delayed in
time and with an acquired time-dependent change in the phase. In this case
the simulated signal $y(t)$ is the sum of an "incident" chirp $y_{0}(t)$ and
a "deformed reflected" chirp $y_{R}(t)$. Noise is added to the signal and the $SNR=10 dB$.
However thanks to the analysis in \cite{Briolle1} we may consider these two
waves separately.

For the "incident" chirp $y_{0}(t)$ the analysis is performed in two
different situations that differ mainly in an amplitude term.

The signal is:

\begin{eqnarray}
y_{0}\left( t\right) &=&b(t),t\in \left[ 0,3\right] s  \notag \\
y_{0}\left( t\right) &=&A(t)e^{i\Phi _{0}(t)}+b(t),t\in \left[ 3,18%
\right] s  \notag \\
y_{0}\left( t\right) &=&b(t),t\in \left[ 18,20\right] s  \label{3.2.1}
\end{eqnarray}
where {$\Phi _{0}(t)=a_{0}t^{2}+b_{0}t$} and $a_{0}$, $%
b_{0}$ are chosen to have $\partial _{t}\phi (3)=75rd/s$ and $\partial
_{t}\phi (18)$=50 rd/s. 

\begin{figure}[tbh]
\begin{center}
\includegraphics[height=6cm,width=9cm]{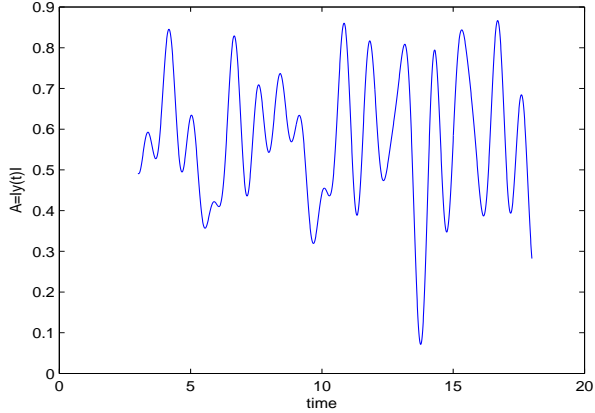}
\end{center}
\caption{Amplitude A(t), defined by equation (\ref{3.1.3}),  of the signal $y(t)$ defined by equation (\ref{3.2.1}).}
\label{Fi:A}
\end{figure}

Here $A(t)$ is $one$ in the first case and in the second case, $A(t)$ defined by (\ref{3.1.3}) is defined for $t\in \left[3,18\right] s $ by equation (\ref{3.1.3}) and presented on Fig.\ref{Fi:A}. Here $N=6$ and $\omega _{k}$ is randomly chosen between $0$ rd/s and $7.5$ rd/s.Notice that for t=14 s, A is very small.

\begin{equation}
A(t)=\frac{\sum_{k=1}^{N}\cos (\omega _{k}t+\phi _{k})+N}{max(A(t))}  \label{3.1.3}
\end{equation}

For this signal, \textbf{tPD} considerably improves the result for $t\in %
\left[ 0,3\right] $ and $t\in %
\left[ 18,20\right] $, as it is easy to understand since the phase derivative
of a random signal may have large local values but the corresponding
amplitude of the total signal be small. 

After using the \textbf{tPD},  we summarize the performances of the different tools in the following Table 1
in terms of their $sdev$\footnote{%
In this case the standard deviation $sdev$ is defined using the difference
between the analytic expression of the known $\partial _{t}\phi (t)$ and the
corresponding estimated value}. \newline

\begin{table}[tbp]
\begin{center}
\begin{tabular}{|c|c|c|c|c|}
\hline\hline
\emph{sdev} & $TDM$ & $TDM+LM_5$ &  $TDM+TBD$ &  $TDM+TBD+LM_5$\\ \hline
$A(t)=1 $ & 38.5 rd/s & 4.5 rd/s & 1.8 rd/s & 1.5 rd/s \\ \hline
$A(t)\neq 1$ & 51 rd/s & 11.2 rd/s & 1.9 rd/s & 1.5 rd/s \\ \hline\hline
 & $TNM$ & $TNM+LM_5$ &  $TNM+TBD$ &  $TNM+TBD+LM_5$\\ \hline
$A(t)=1 $ & 23.5 rd/s & 3.9 rd/s & 1.8 rd/s & 1.5 rd/s \\ \hline
$A(t)\neq 1$ & 39.3 rd/s & 10.7 rd/s & 1.3 rd/s & 2.2 rd/s \\ \hline\hline
\end{tabular}%
\end{center}
\caption{ Comparison of the different tools in terms of their $sdev$
for the signal $y_0$ defined by (\protect\ref{3.2.1}).}
\label{tab:3.2.1}
\end{table}

\begin{figure}[htb]
\begin{center}
\includegraphics[height=6cm,width=9cm]{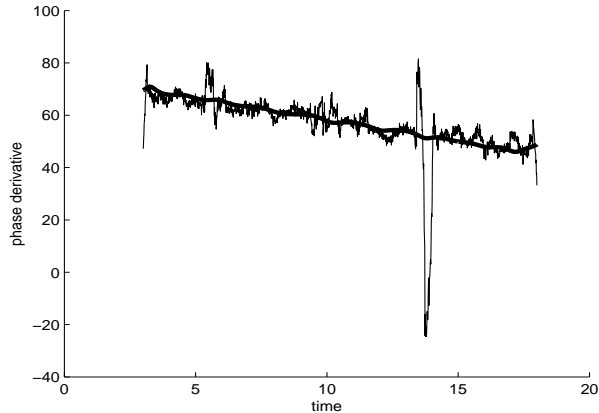}
\end{center}
\caption{Phase derivative of $y_0(t)$, defined by equation(\protect\ref%
{3.2.1}) for the case $A(t)\neq 1$, using \textbf{TDM+LM} ($LM_5$) and \textbf{TDM+TBD} (bold line) on the tomogram for $\theta=\frac{\pi}{5}$.}

\label{Fi:fchirp-filtTDM-TBD-pi_5}
\end{figure}


In Fig.\ref{Fi:fchirp-filtTDM-TBD-pi_5} we show the graphic representation of the
reconstructed phase derivative, for $A(t)\neq 1$, corresponding to \textbf{TDM+LM} ($LM_5$) and \textbf{TDM+TBD} of Table 1. As
can be seen, the combined use of the tools described in section 2 allows a
very efficient reconstruction of the phase derivative in this case, except when the signal is very small, for $t \approx 14s$. In particular \textbf{TDM} (or \textbf{TNM} )+ \textbf{TBD} gives very good results for an amplitude varying
signal. It is however worthwhile to mention that the tomogram spectral
family (\ref{2.2}) is particularly well adapted to this type of "incident
wave" since in the limit case of an infinite time domain the corresponding
spectrum would be reduced to a unique $c_{x}^{\theta }(f)$. But if, on one
hand, we take advantage of this fact because the incident wave in
reflectometry has this shape, on the other hand, the next example shows
that the good performances of the tool are not limited to this particular
non linear phase shape.

We also notice from Table 1 that, even before filtering, the normalisation \textbf{TNM} improves the results. This arises mostly from the processing of the small amplitudes regions.

Let now consider the "deformed reflected chirp" $y_R(t)$ defined by :

\begin{equation}
y_{R}(t)=A(t)e^{i\Phi _{R}(t)}+b(t),t\in \left[ 0,20\right] s  \label{3.2.2}
\end{equation}
where {$\Phi _{R}(t)=a_{R}t^{2}+b_{R}t+10t^{\frac{3}{2}}$} and $a_{R}$, $%
b_{R}$ are chosen to have $\partial _{t}\phi (0)=75rd/s$ and $\partial
_{t}\phi (20)$=50 rd/s. 
As before $A(t)$ is $one$ in the first case and
defined by (\ref{3.1.3}) in the second case with, in each case, a noisy
component $b(t)$ with $SNR=10dB$.

The performances of the different tools, summarised in Table 2 in terms of their $sdev$,
show how the tomogram based tools perform very accurate
estimations of the (local) phase derivative in cases where other methods may
have some difficulties. In particular, \textbf{TBD} seems a very efficient
method to denoise the signal as it can be seen in the Fig.\ref%
{Fi:yR-filtTDM-TBD-pi_5}.

\begin{table}[tbp]
\begin{center}
\begin{tabular}{|c|c|c|c|c|}
\hline\hline
\emph{sdev} & $TDM$ & $TDM+LM_5$ &  $TBD$ &  $TBD+LM_5$\\ \hline
$A(t)=1 $ & 27.1 rd/s & 6.6 rd/s & 2.0 rd/s & 1.5 rd/s \\ \hline
$A(t)\neq 1$ & 92 rd/s & 24.7 rd/s & 3.0 rd/s & 2.9 rd/s \\ \hline\hline
 & $TNM$ & $TNM+LM_5$ &  $TNM+TBD$ &  $TNM+TBD+LM_5$\\ \hline
$A(t)=1 $ & 27.1 rd/s & 4.7 rd/s & 2.0 rd/s & 1.5 rd/s \\ \hline
$A(t)\neq 1$ & 36.1 rd/s & 14.1 rd/s & 2.2 rd/s & 1.4 rd/s \\ \hline\hline
\end{tabular}%
\end{center}\caption{Comparison of the different tools in terms of their $sdev$
for the signal $y_R$ defined by (\protect\ref{3.2.2}).}
\label{tab:3.2.2}
\end{table}


\begin{figure}[htb]
\begin{center}
\includegraphics[height=6cm,width=9cm]{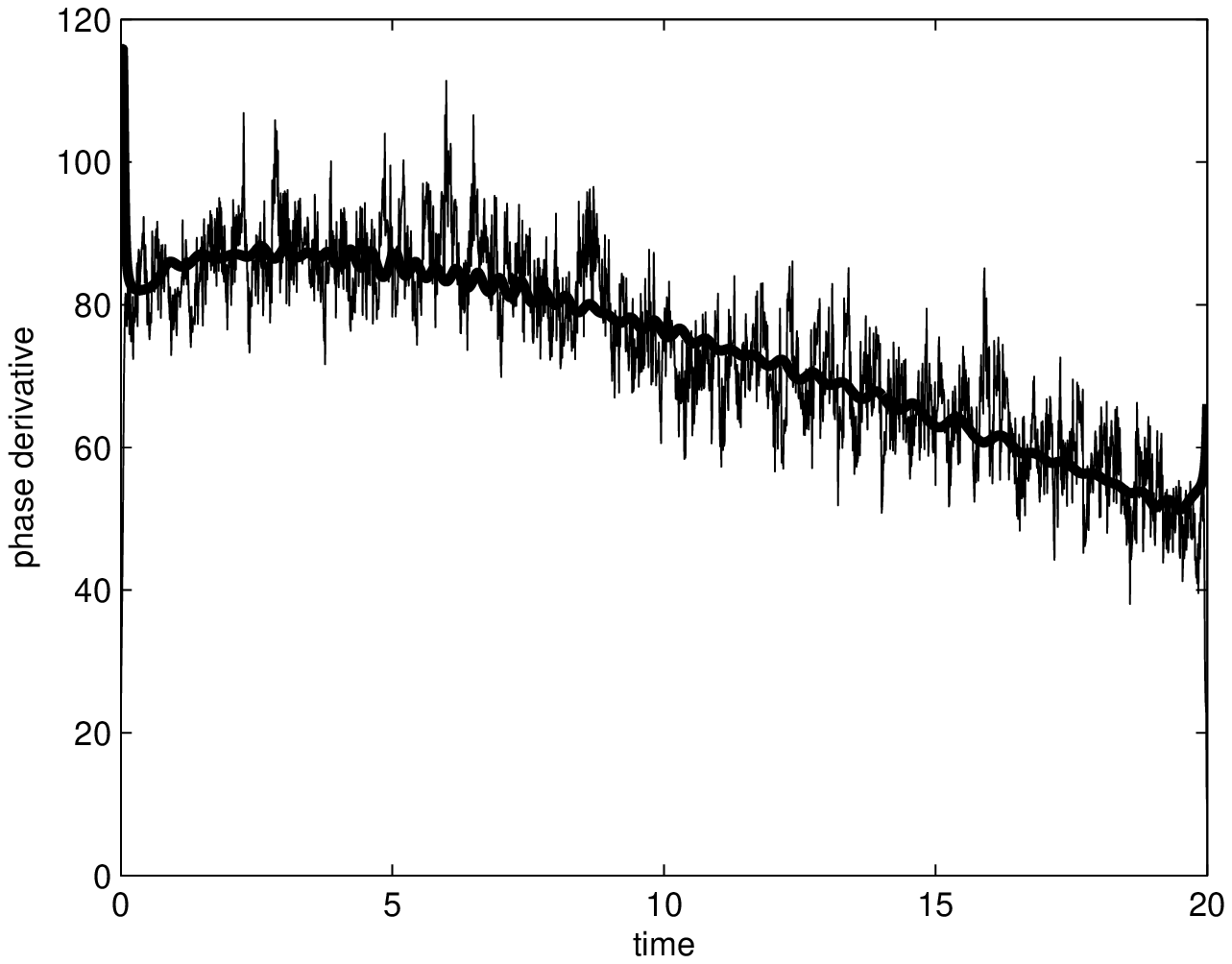}
\end{center} 
\caption{Phase derivative of $y_R(t)$, defined by equation(\protect\ref%
{3.2.2}) for the case $A(t)\neq 1$, using \textbf{TDM+LM} ($LM_5$) and \textbf{TDM+TBD} (bold line) on the tomogram for $%
\protect\theta=\frac{\protect\pi}{5}$.}
\label{Fi:yR-filtTDM-TBD-pi_5}
\end{figure}


\section{Application to reflectometry data}

We now show the ability of the tomographic methods to extract the phase
derivative of an experimental signal coming from reflectometry measurements
during a discharge in the Tore Supra at Cadarache.

\begin{figure}[htb]
\begin{center}
\includegraphics[height=6cm,width=9cm]{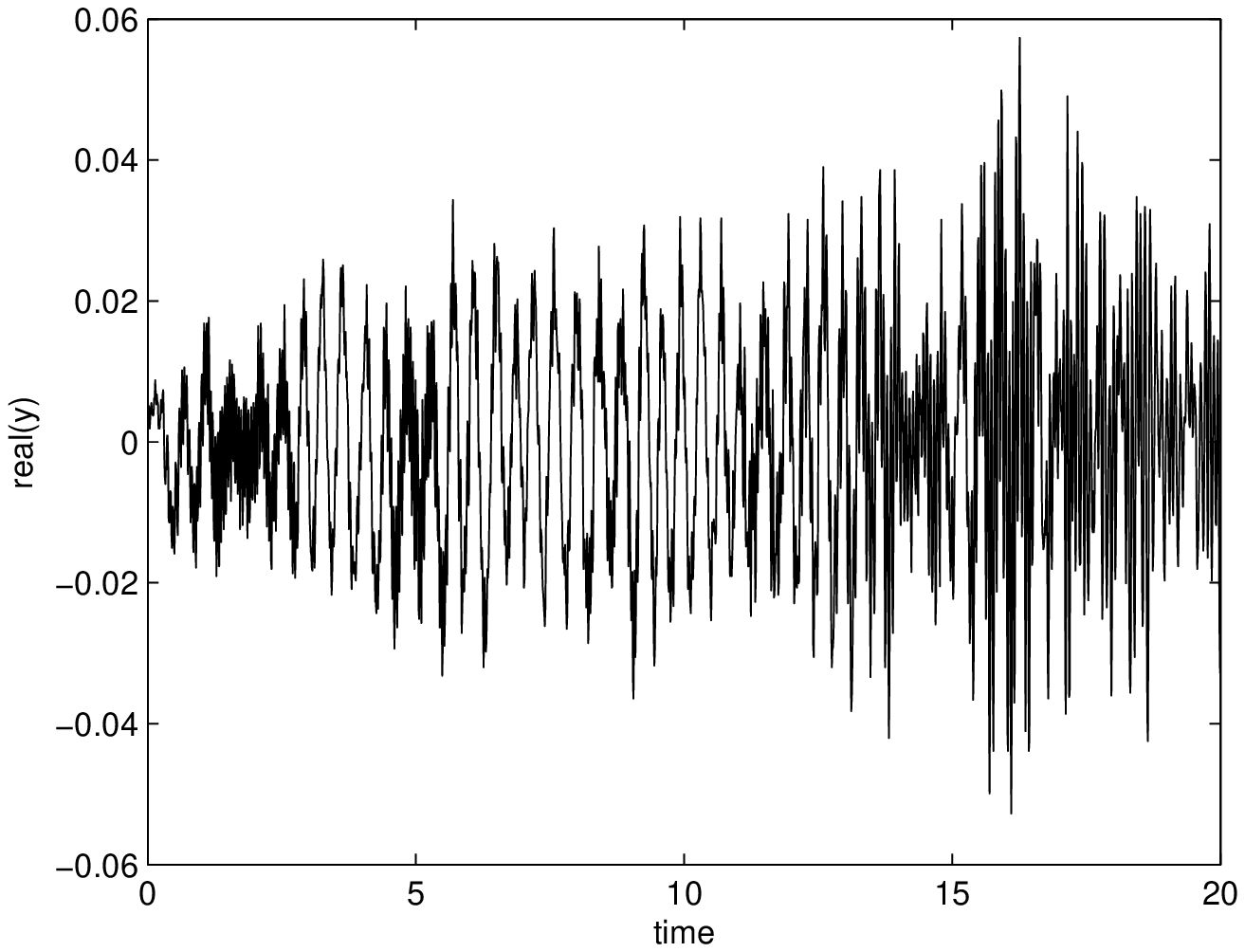}
\end{center}
\caption{Time representation of the reflectometry signal (real part). }
\label{Fi:reflec-y}
\end{figure}

The sweep-frequency reflectometry system of Tore Supra launches a probing
wave on the extraordinary mode polarization (X mode) in the V band (50--75
GHz)~\cite{Clairet},~\cite{Clairet2}, ~\cite{Clairet3}. The emitting and
receiving antennas are located at about 1.20 m from the plasma edge, outside
the vacuum vessel. The reflectometry system repeatedly sends sweeps of
duration $20\mu s$. The heterodyne reflectometers, with $I/Q$ detection,
provide a good Signal to Noise Ratio, up to $40dB$. For each sweep, the
reflected chirp $E_{R}(t)$ is mixed with the incident sweep $E_{0}(t)$ and
only the interference term is recorded as an in-phase and a $90^{\circ }$
phase shifted sampled signals: 
\begin{equation*}
x_{1}(t)=A_{0}A_{R}(t)\cos (\varphi (t)) 
\end{equation*}
\begin{equation*}
x_{2}(t)=A_{0}A_{R}(t)\sin (\varphi (t)) 
\end{equation*}
Let the reflected signal be

\begin{equation}
y(t)=x_{1}(t)+ix_{2}(t)=A(t)e^{i\varphi (t)}  \label{4.1}
\end{equation}
The phase derivative of the signal corresponding to the plasma component of $%
y(t)$ is used to localize the cut-off density in the plasma. The amplitude
of this signal $A(t)=A_{0}A_{R}(t)$ corresponds to a low frequency. The real
part of the signal $y(t)$ is shown in Fig.\ref{Fi:reflec-y}.

The tomogram at $\theta =\frac{3\pi }{10}$ was used to perform the
factorization of the signal in \cite{Briolle1}. Cutting the spectrum at $%
\epsilon =0.05$ max($\left\vert c_{x_{n}}^{\theta }(y)\right\vert $) the
signal is factorized in three components as shown on Fig.\ref%
{Fi:reflec-cn-3pi_10-partial}.

\begin{figure}[htb]
\begin{center}
\includegraphics[height=5cm,width=9cm]{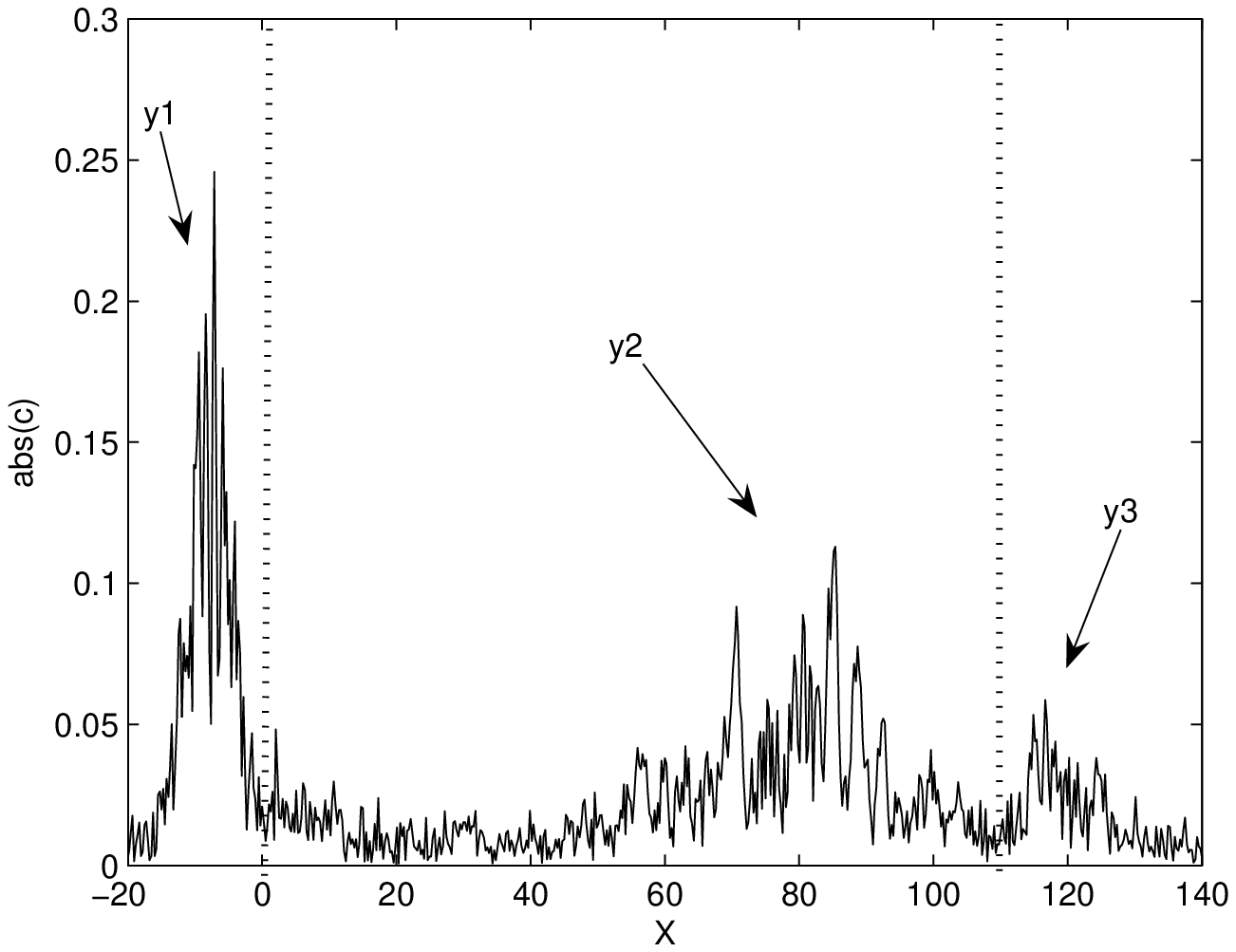}
\end{center}
\caption{Spectrum $c^{\protect\theta}_{x_n}(y)$ of the reflectometry signal $%
y(t)$ for $\protect\theta_0=\frac{3\protect\pi}{10}$ used in the
factorization.}
\label{Fi:reflec-cn-3pi_10-partial}
\end{figure}


\subsection{First component, the reflection on the porthole}

The first component, $y_{1}(t)$, of the reflectometry signal is a low
frequency signal corresponding to the heterodyne product of the probe signal
with the reflection on the porthole \cite{Clairet3}. This complex signal is
written :

\begin{equation}
y_{1}(t)=A_{1}(t)e^{i\varphi _{1}(t)}  \label{4.3}
\end{equation}

The phase derivative $\partial _{t}\varphi _{1}(t)$ may be positive and
proportional to the time $\tau _{1}$ of the reflection of the probe signal
on the porthole. If not, the reflectometry signal $y(t)$ defined by (\ref%
{4.1}) is multiplied by $e^{iat}$ for some $a$ to calibrate the measurement.
The real part and the modulus of $y_{1}(t)$ are shown in Fig.\ref%
{Fi:reflec-absy1-3pi_10}.

\begin{figure}[htb]
\begin{center}
\includegraphics[height=5cm,width=9cm]{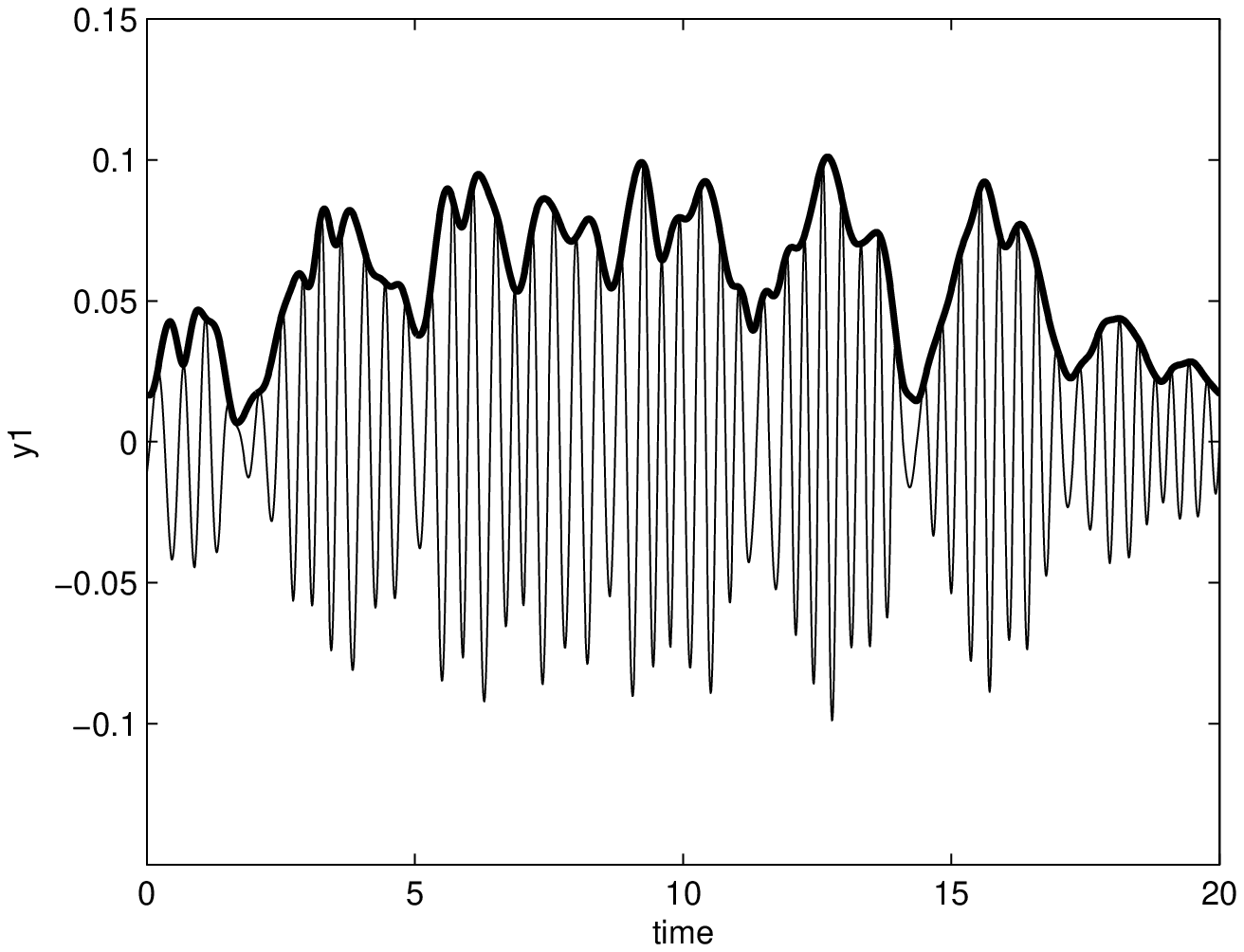}
\end{center}
\caption{First component of the reflectometry signal corresponding to the
reflection on the porthole (modulus and real part).}
\label{Fi:reflec-absy1-3pi_10}
\end{figure}

\begin{figure}[htb]
\begin{center}
\includegraphics[height=5cm,width=9cm]{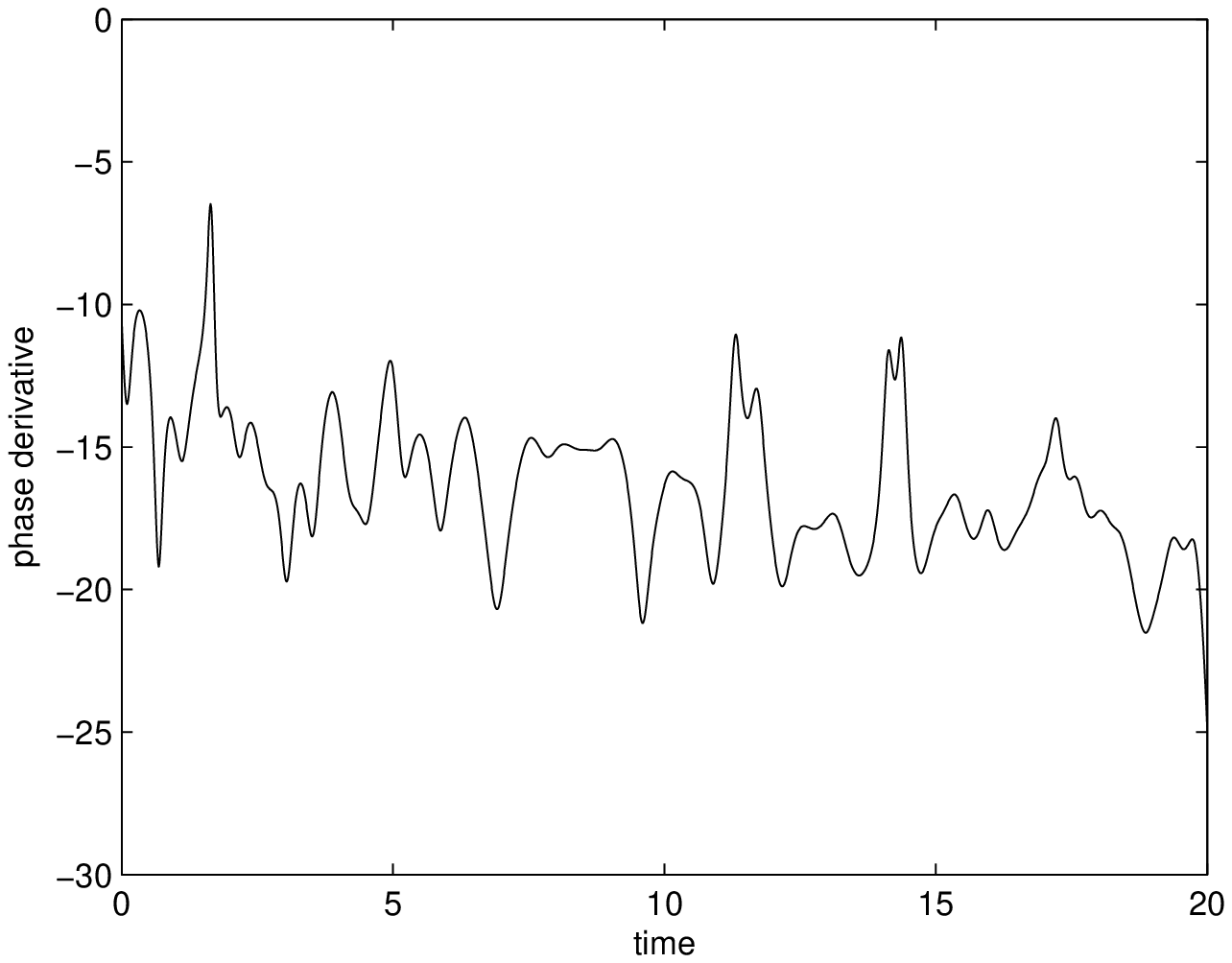}
\end{center}
\caption{Phase derivative of the first component of the reflectometry signal
corresponding to the reflection on the porthole, using \textbf{TDM} 
on the tomogram
for $\protect\theta=\frac{3\protect\pi}{10}$.}
\label{Fi:reflec-phasey1-3pi_10}
\end{figure}

The phase derivative of $y_{1}(t)$ was then computed using \textbf{TDM} and
shown in Fig.\ref{Fi:reflec-phasey1-3pi_10}. 
The mean value of the
phase derivative is equal to $-16.4$ $rd/s$ and its sdev to $2.4rd/s$ (less
than $15$ $\%$) in agreement with a rough estimation based on a spectrogram
technique (see section 5). Since the phase derivative of $y_{1}(t)$ is
negative, the reflectometry signal $y(t)$ had to be calibrated to set $%
\partial _{t}\varphi _{1}(t)$ proportional to $\tau _{1}$ (see conclusion).

We also shift the phase derivative of the other components by the same value.

\subsection{Second component, the plasma signal}

The second component has a Fourier spectra that fits the expected behavior
corresponding to the reflection of the wave inside the plasma of the Tore
Supra \cite{Clairet3}. This component, $y_{2}(t)$ is defined as:

\begin{equation}
y_{2}(t)=A_{2}(t)e^{i\varphi _{2}(t)}  \label{4.4}
\end{equation}
The modulus and real part are displayed together in the same plot (Fig.\ref%
{Fi:reflec-absy2-3pi_10}). 

\begin{figure}[tbh]
\begin{center}
\includegraphics[height=5cm,width=9cm]{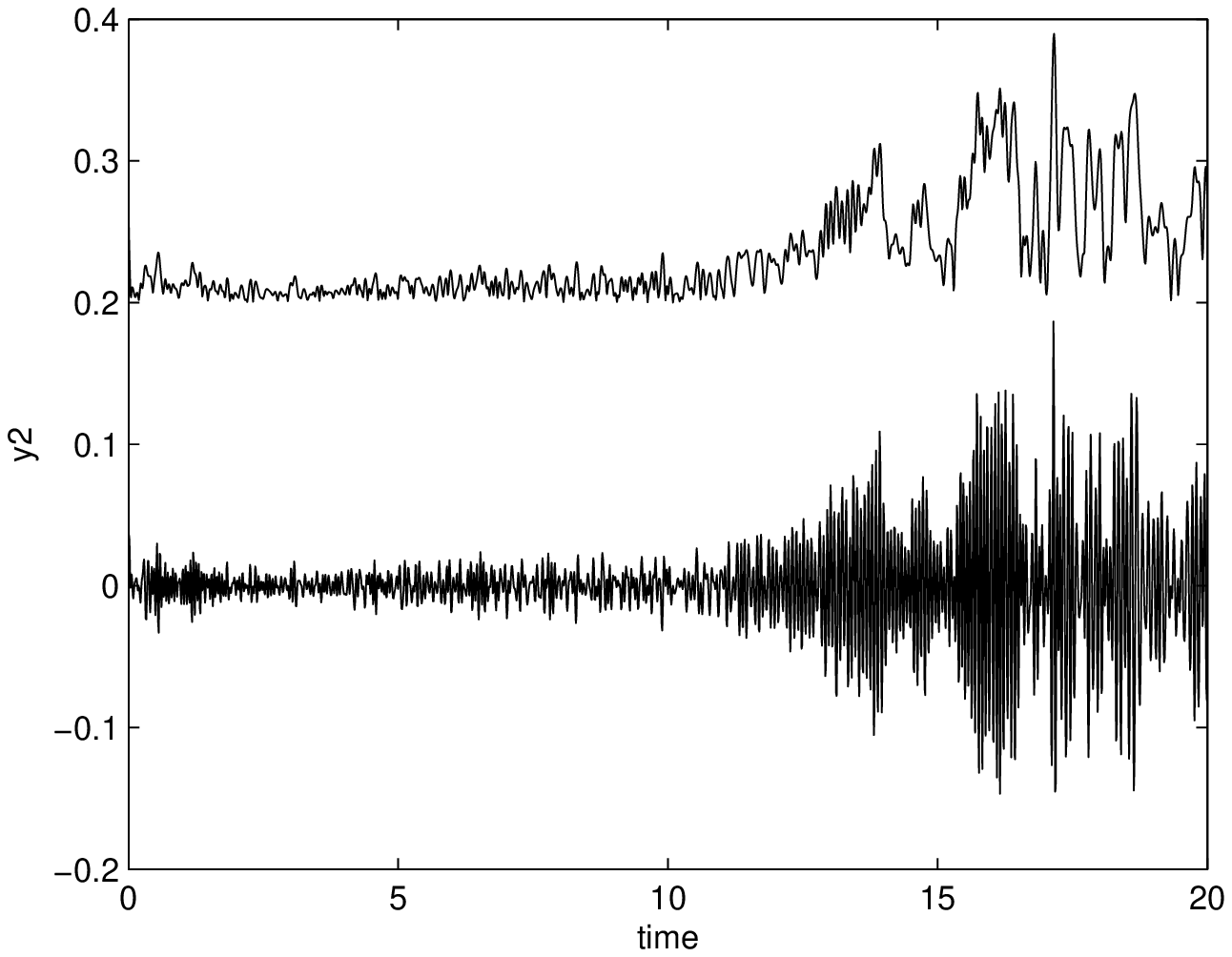}
\end{center}
\caption{Modulus and real part of the second component of the reflectometry
signal, corresponding to the reflection on the plasma. For visual puposes,
the average of the modulus is shifted by 0.2.}
\label{Fi:reflec-absy2-3pi_10}
\end{figure}

Even if the modulus is of low frequency in comparison to the real part of $%
y_{2}(t)$, the modulus is not constant. In particular the signal is very
small in the first half. 

The amplitude of the signal $y_{2}$, for $t\in \lbrack 0,11]s$, is very
small in comparison to the amplitude for $t\in \lbrack 11,20]s$. The  power ratio 
of the signal $y_{2}$ for $t\in \lbrack 0,11]s$ to the one for $%
t\in \lbrack 11,20]s$ equals $0.035$ \footnote{%
The power $P_{s}$ of a signal $s(t)$ is defined by :
\par
\begin{equation}
P_{s}=\frac{1}{T}\int_{0}^{T}\left\vert s(t)\right\vert ^{2}dt  \notag
\end{equation}%
} (SNR $\approx-15$ dB). Then the first part of the signal can be
considered as filtered noise and , with \textbf{tPD}, the phase derivative will not be computed.
This fact may correspond to the difficulty of
the incident wave to reach the plasma in the first (lower) band of
"instantaneous frequencies" and therefore to a bad accuracy in the detection
of the low densities present at the border of the plasma in tokamaks.

\begin{figure}[htb]
\begin{center}
\includegraphics[height=5cm,width=9cm]{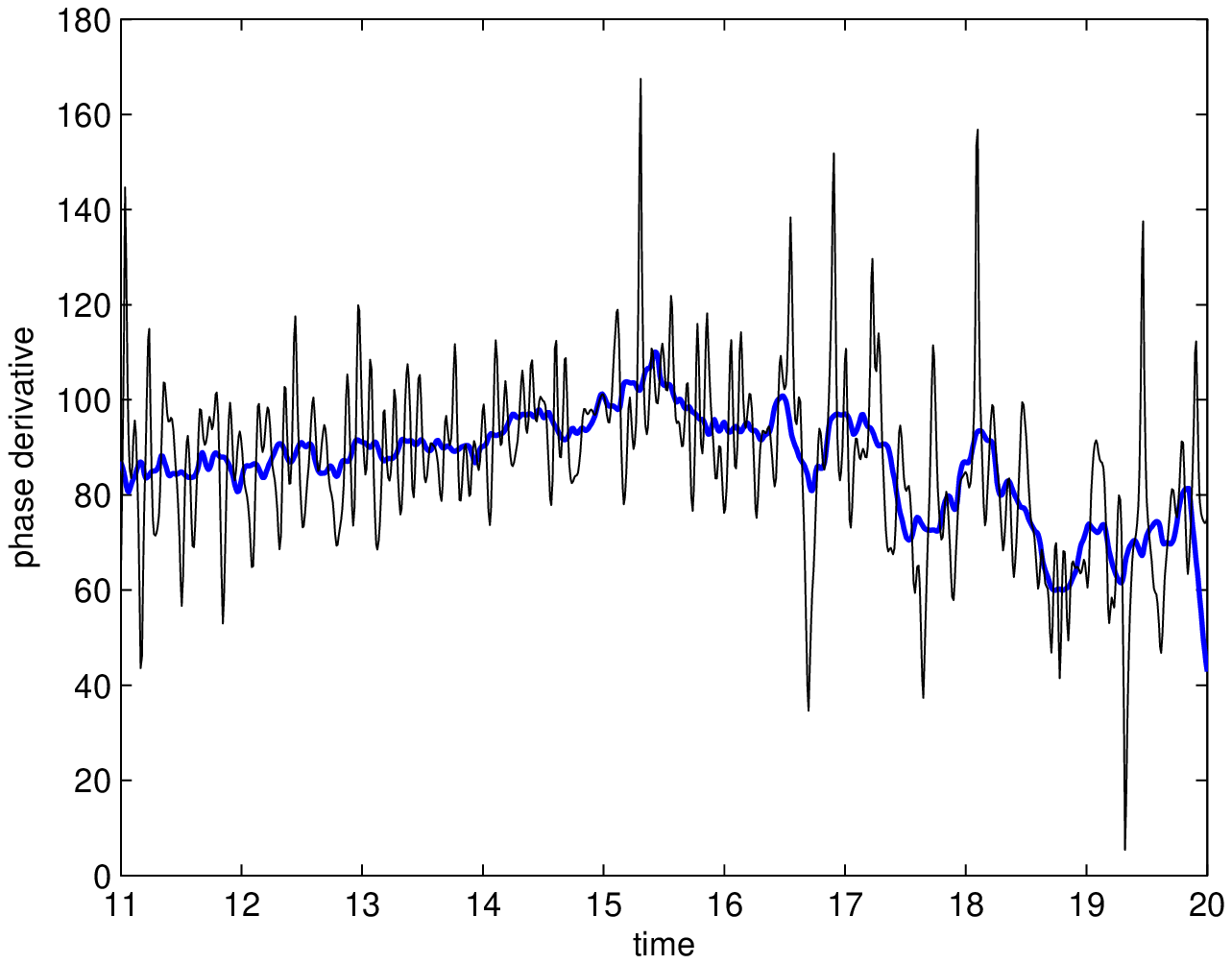}
\end{center}
\caption{Phase derivative of the second component of the reflectometry
signal for $t\in [11,20]s$. \textbf{TDM} and the \textbf{TDM+LM} filtered
phase derivative ($LM_{15}$ : bold) are presented on the same plot.}
\label{Fi:reflec-phasey2-3pi_10}
\end{figure}

The phase derivative of the last part of the signal, for $t\in \lbrack
11,20]s$, corresponding to  \textbf{TDM} and  \textbf{TDM+LM}
 are shown in Fig.\ref{Fi:reflec-phasey2-3pi_10}.

\begin{figure}[htb]
\begin{center}
\includegraphics[height=5cm,width=9cm]{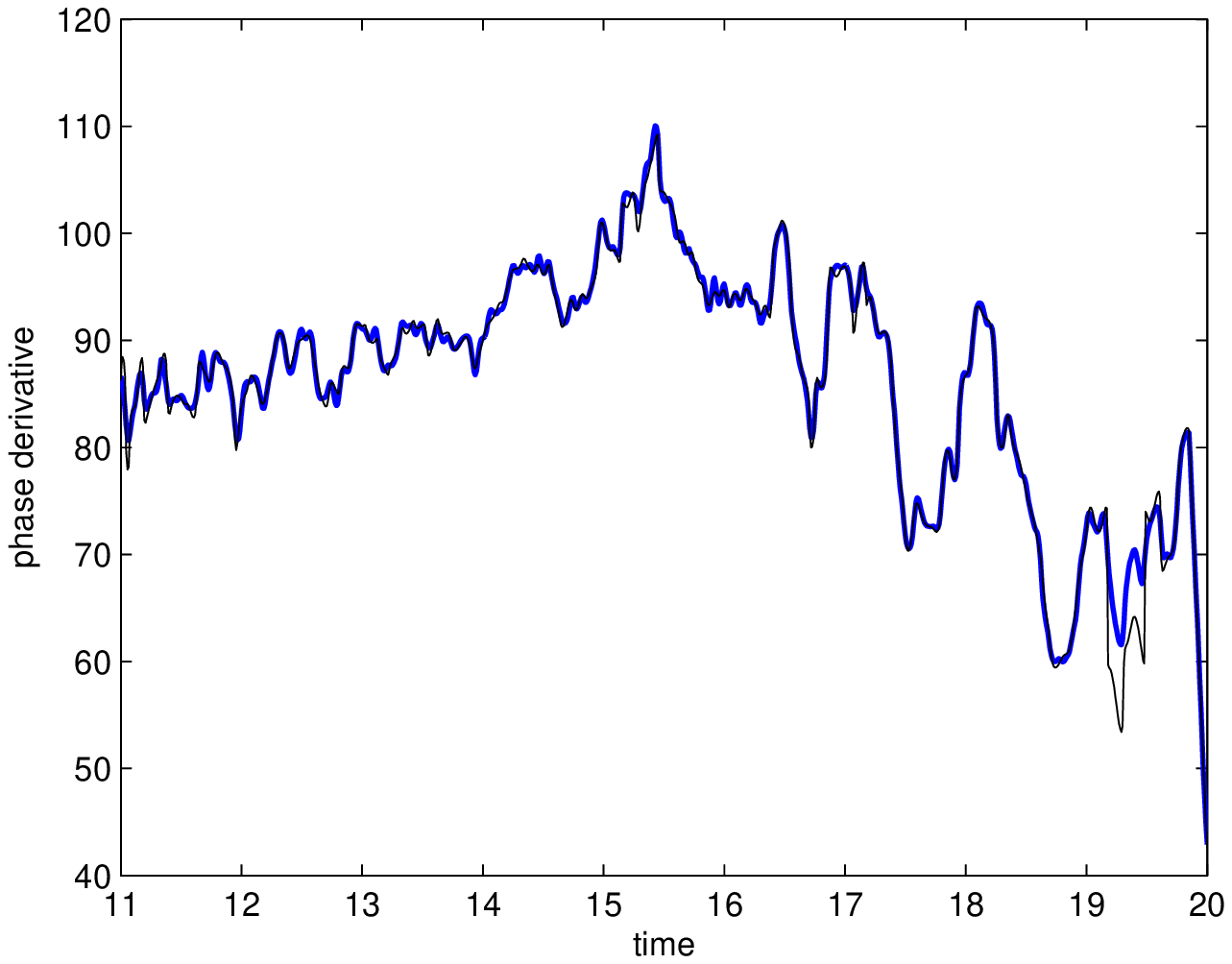}
\end{center}
\caption{For $t\in [11,20]s$, \textbf{$LM_{15}$} filtered phase derivative
of the second component of the reflectometry signal obtained with \textbf{TDM%
} (thin line) and \textbf{TNM} (bold line).}
\label{Fi:reflec-phasey2-filt-3pi_10}
\end{figure}

For comparison, the \textbf{$LM_{15}$} filtered phase derivative of the 
\textbf{TDM} and of the \textbf{TNM} (bold), for $t\in \lbrack 11,20]s$, are
plotted on the same Fig.\ref{Fi:reflec-phasey2-filt-3pi_10}. The results are
nearly the same, except for small differences for $t\approx 19.5s$ where the
amplitude of the signal is very small.

We conclude that combining \textbf{TDM} (or \textbf{TNM})  with \textbf{LM}
gives an accurate estimation of the phase derivative of the plasma
component. \textbf{TNM} appears to be performant when  the amplitude of the signal small. This claim
will be confirmed by a comparison with the usual spectrogram analysis in
section 5.


\subsection{Third component, the multireflection}

The last component of the reflectometry signal corresponds, \cite{Clairet3} 
\cite{Briolle1}, to multireflections of the wave on the wall of the vacuum
vessel. This component, $y_{3}(t)$ is written as :

\begin{equation}
y_{3}(t)=A_{3}(t)e^{i\varphi _{3}(t)}  \label{4.5}
\end{equation}

\begin{figure}[htb]
\begin{center}
\includegraphics[height=5cm,width=9cm]{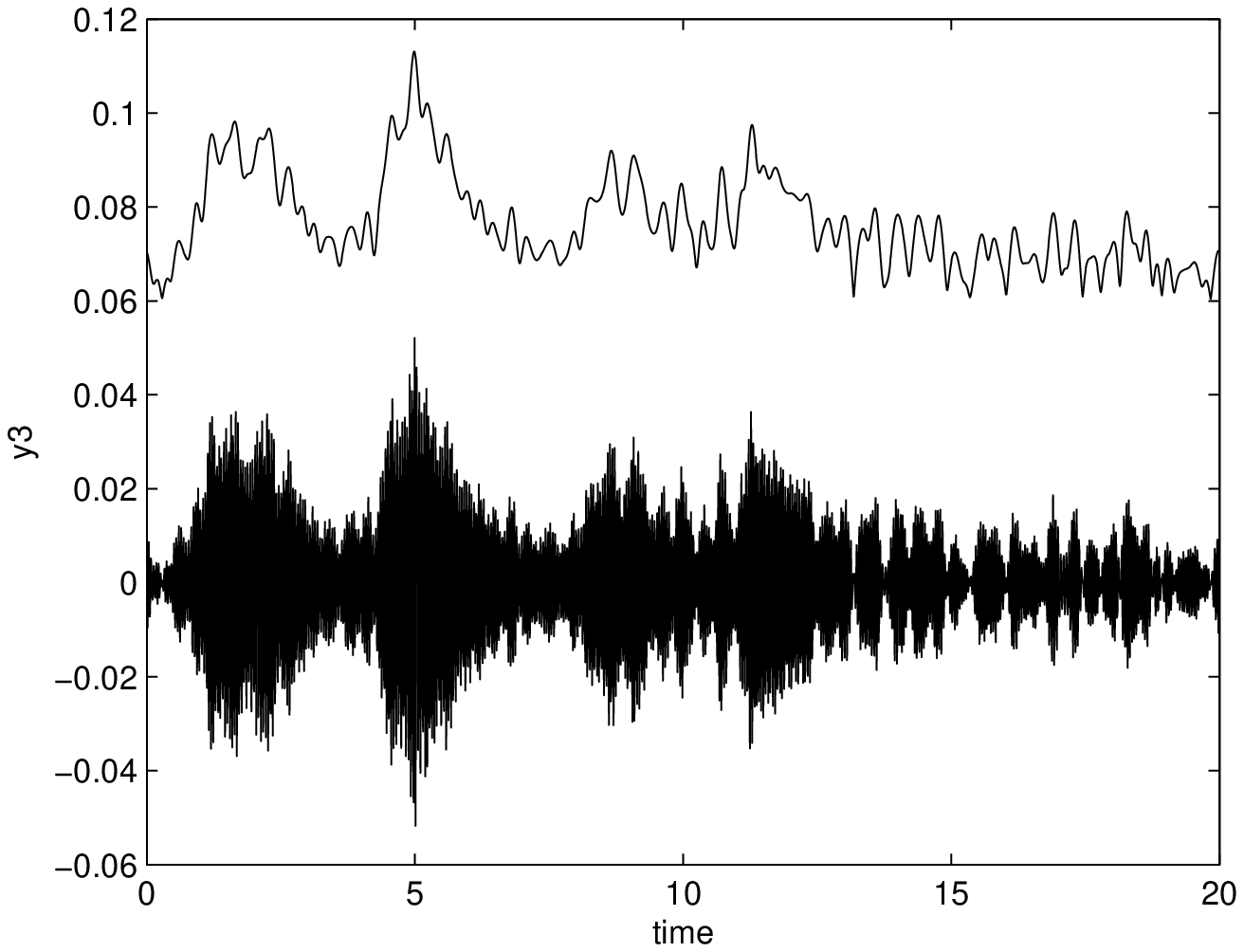}
\end{center}
\caption{Modulus (bold) and real part of the third component of the reflectometry
signal, corresponding to multi reflections on the vessel. For visual
puposes, the average of the modulus is shifted by 0.06.}
\label{Fi:reflec-absy3-3pi_10}
\end{figure}

The modulus $A_{3}(t)$ and the real part of $y_{3}(t)$ are presented
together on the same figure (Fig.\ref{Fi:reflec-absy3-3pi_10}).  As compared to the
real part of $y_{3}(t)$, the modulus $A_{3}(t)$ is a low frequency signal.
We notice that for $t>13s$ the modulus is very small.

\begin{figure}[htb]
\begin{center}
\includegraphics[height=5cm,width=9cm]{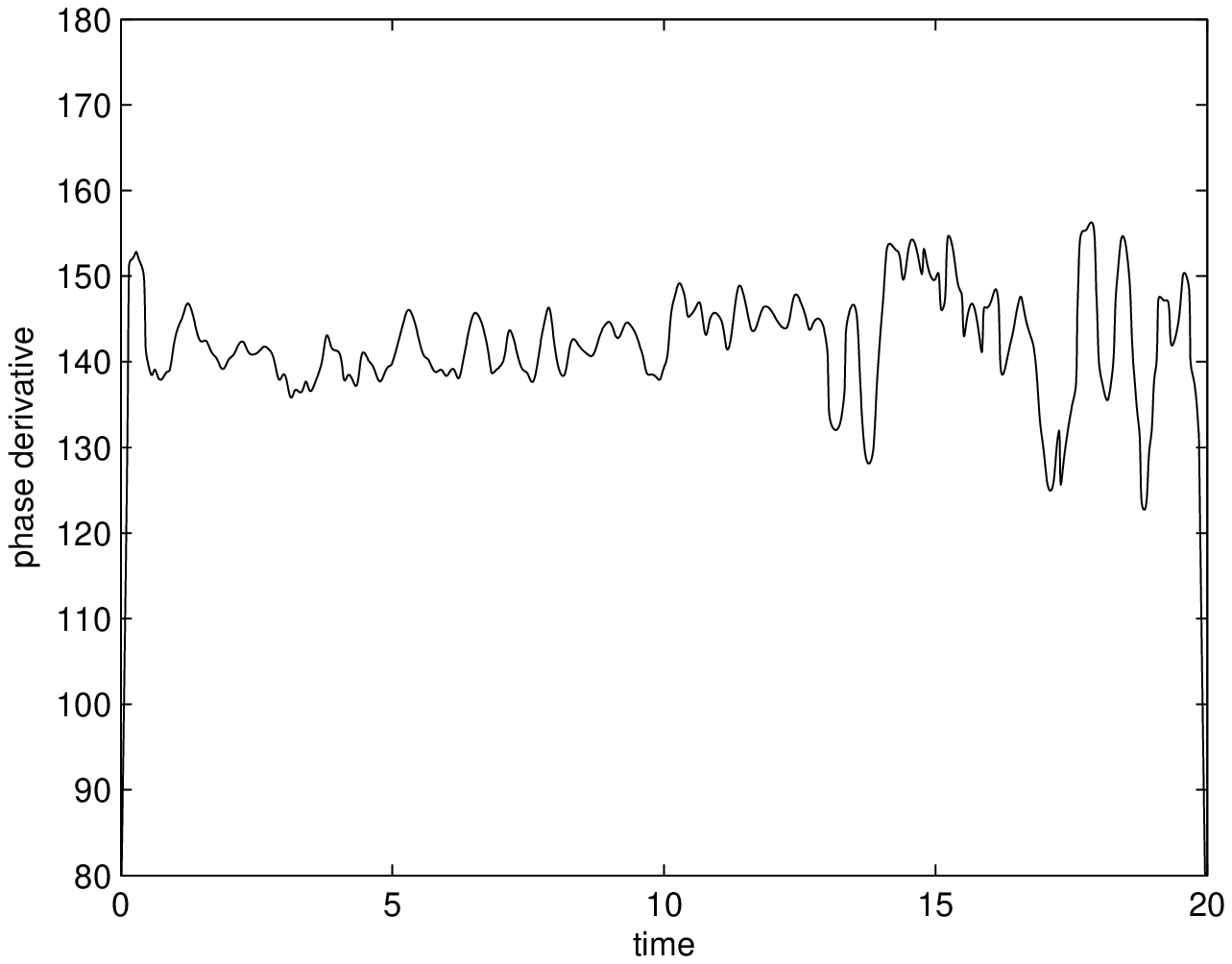}
\end{center}
\caption{Phase derivative of the third component of the reflectometry
signal, corresponding to multi reflections on the vessel, using \textbf{TDM+LM}
($LM_{15}$)on the tomogram for $\protect\theta=\frac{3\protect\pi}{10}$.}
\label{Fi:reflec-phasey3-filt-3pi_10}
\end{figure}

The phase derivative of $y_{3}(t)$ estimated using \textbf{TDM+LM} is plotted in
Fig.\ref{Fi:reflec-phasey3-filt-3pi_10}.

\subsection{Components comparison}

The \textbf{LM} filtered phase derivative of the three components of the
reflectometry signal are plotted together on the same figure (Fig.\ref%
{Fi:reflec-phasey1y2y3-filt-3pi_10}). It is instructive to compare these
phase derivatives. For the first component, the phase derivative $\partial
_{t}\varphi _{1}(t)$ is almost constant. This is because the phase $\varphi
_{1}(t)$ is mainly due to a simple reflection on a nearby object, the
porthole. The reflection on the plasma is quite complex. The first part of
the signal should be considered as filtered noise and shows that there is a
problem in reconstructing the density profile corresponding to this part of
the sweep. The phase derivative of the third component of the signal,
corresponding to multi reflections on the vessel, presents some similarities
with the phase derivative of the first component, except for $t>14.5s$. It
will eventually be interesting to factorize again this component if some
information related to the properties of the plasma close to the vessel
walls can be extracted from it. The modulus of the three components are low
frequency signals, compared to the signals themselves. Usually, the phase
derivative obtained by \textbf{TDM} is accurate when filtered by \textbf{LM}%
. In this case \textbf{TBD} does not seem adequate for denoising purposes,
because it correlates with the component analysis and may eventually corrupt
the factorization of the signal.

\begin{figure}[tbh]
\begin{center}
\includegraphics[height=5cm,width=9cm]{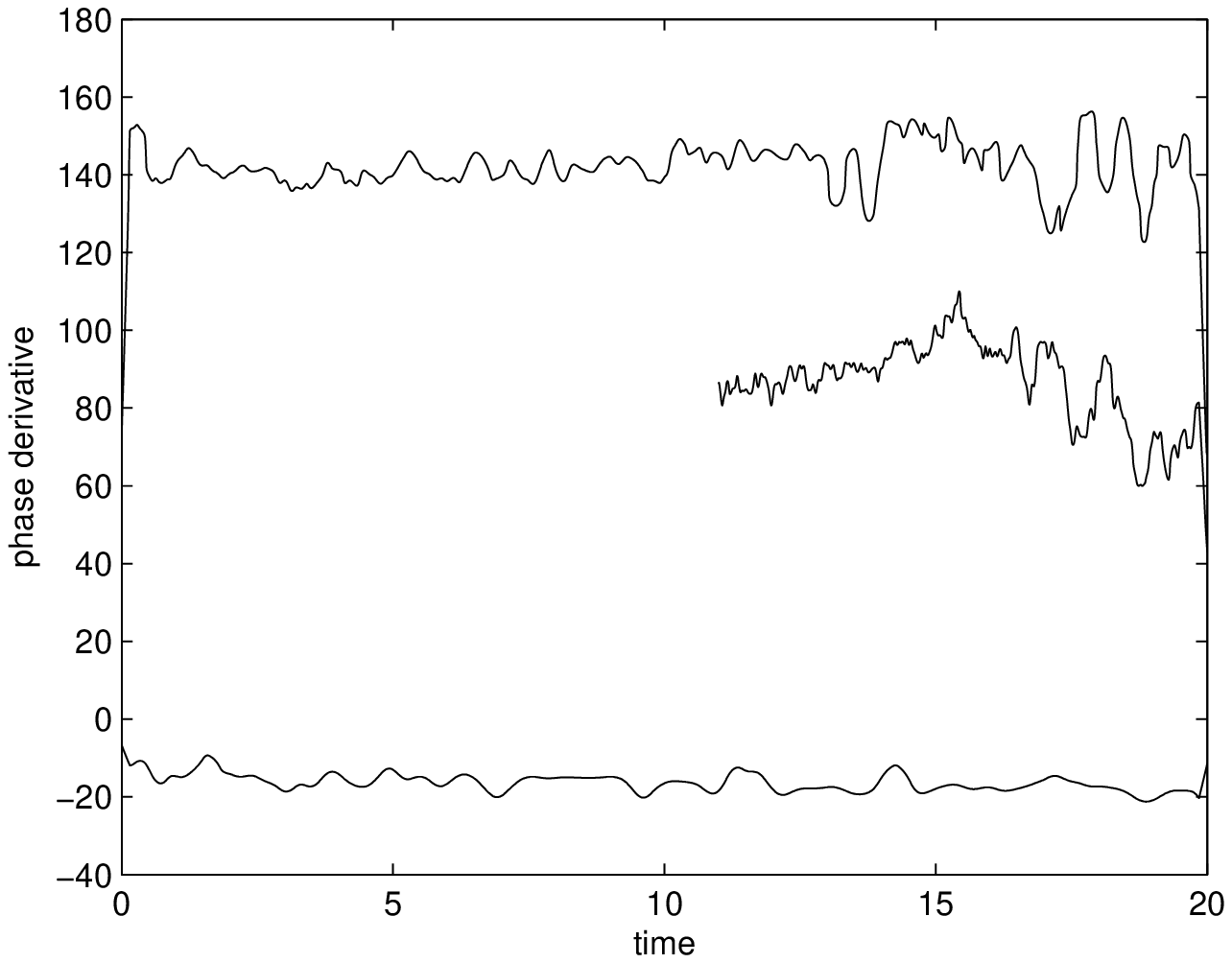}
\end{center}
\caption{Phase derivative of the three
components of the reflectometry signal, obtained with \textbf{TDM+LM} ($LM_{15}$). The mean value of $\partial_t 
\protect\varphi_1(t)$ is equal to -18 $rd/s$, $\partial t \protect\varphi%
_2(t) $ to 75 $rd/s$ and of $\partial t \protect\varphi_3(t) $ to 140 $rd/s$%
. }
\label{Fi:reflec-phasey1y2y3-filt-3pi_10}
\end{figure}


\section{Tomograms and spectrogram analysis}


In this section we obtain the "frequency" of the signal as a function of
time, obtained by a moving window FFT spectrogram, and compare it with the
phase derivative obtained by the tomographic techniques described before.
The spectrogram is computed with a 64 points length window (the $grain$) and
a $75\%$ overlap rate using the maximum pick method \cite{Boashash},
allowing a FFT resolution of $10rd/s$ on $121$ time points. We avoided an
estimation with higher time resolution because of FFT resolution
constraints. For the tomographic phase derivative estimation we used, as
usually, \textbf{TDM} (or \textbf{TNM})  together with \textbf{LM}
filtering.

\begin{figure}[htb]
\begin{center}
\includegraphics[height=5cm,width=9cm]{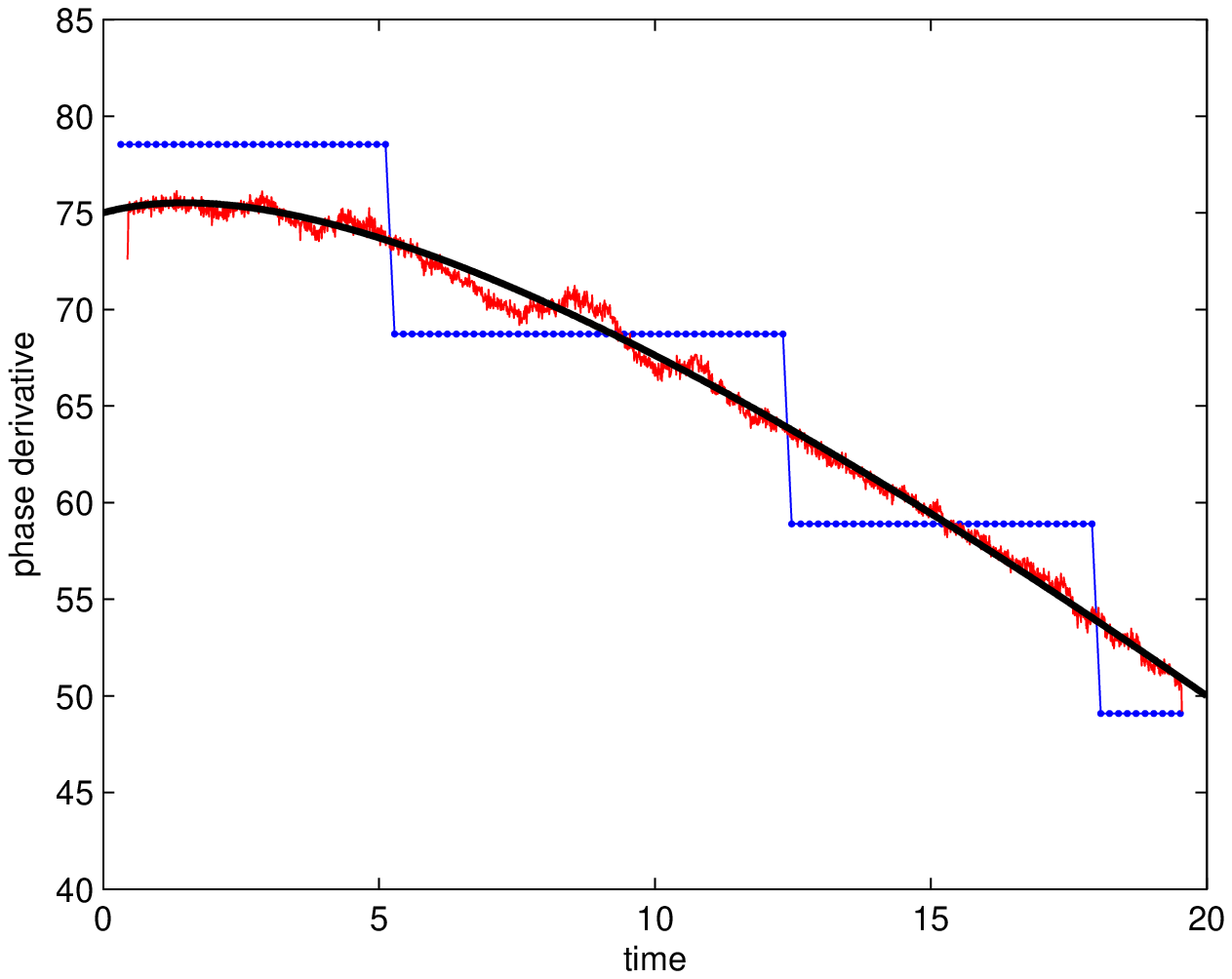}
\end{center}
\caption{Phase derivative of $y_R(t)$ defined by (\protect\ref{3.2.2}) using 
\textbf{TDM+TBD} on the tomogram for $\protect\theta=\frac{\protect\pi}{5}$. The
bold line is the analytical curve of the phase derivative. For comparison
the dotted lines correspond to the maxima of a moving window FFT spectrogram with a resolution $%
\pm 9.9 rd/s$ (see section 5). }
\label{Fi:chirpR-15dB}
\end{figure}

For the simulated "deformed reflected chirp" $y_R(t)$ (Eq \ref{3.2.2}), figure \ref{Fi:chirpR-15dB} shows that a tomogram based
technique gives a much better agreement with the known analytical phase
derivative.

\begin{figure}[htb]
\begin{center}
\includegraphics[height=5cm,width=9cm]{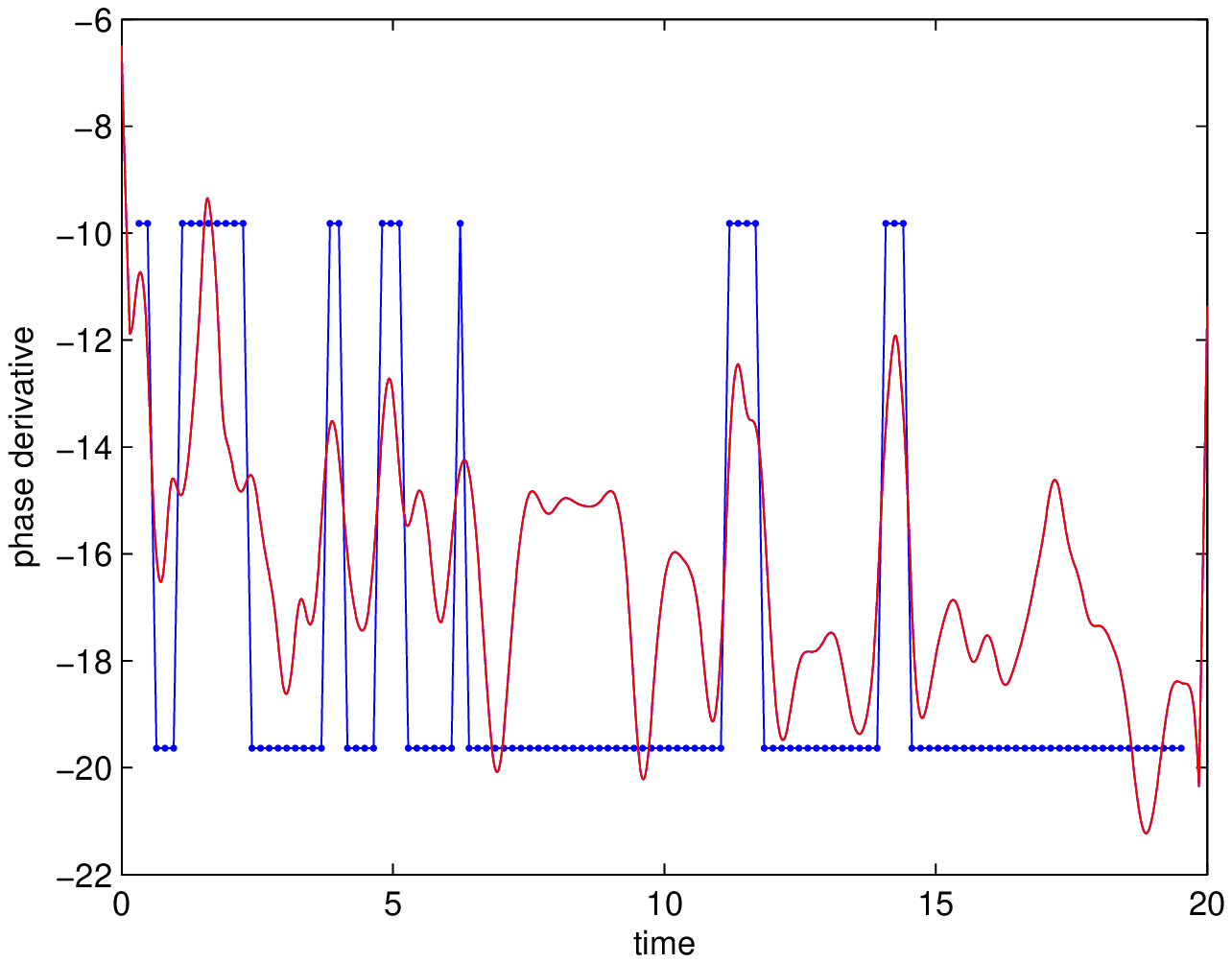}
\end{center}
\caption{For the first component of the reflectometry signal, corresponding
to the reflexion on the porthole phase
derivative obtained by \textbf{TDM+LM} ($LM_{15}$). The dots are the maximum of a moving
window FFT spectrogram with a resolution of $\pm 9.9 rd/s$ .}
\label{PD-y1}
\end{figure}

\begin{figure}[htb]
\begin{center}
\includegraphics[height=5cm,width=9cm]{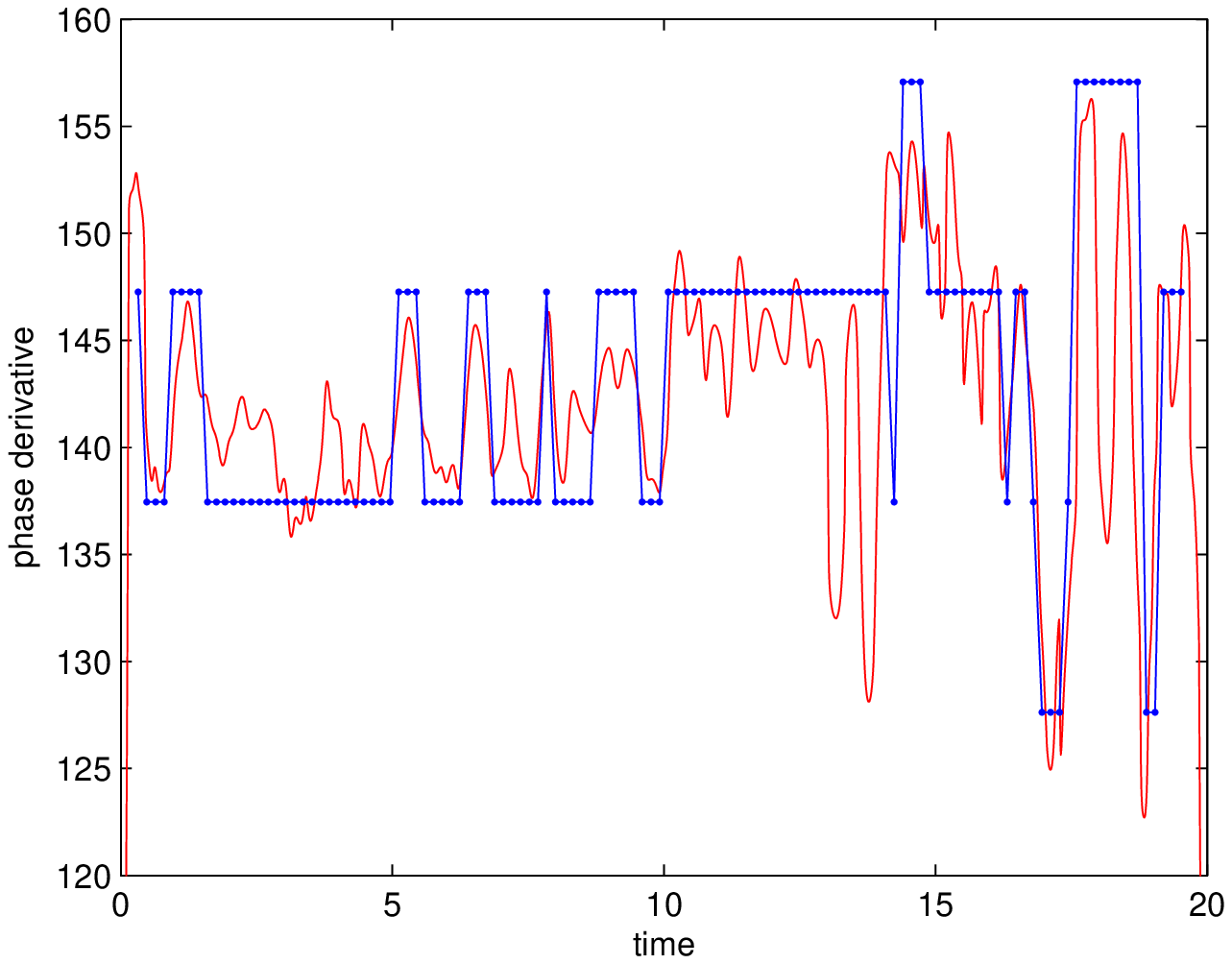}
\end{center}
\caption{For the third component of the reflectometry signal, corresponding
to multi reflexion phase derivative obtained by \textbf{TDM+LM} ($LM_{15}$). The dots are the maximum of a moving window FFT spectrogram with a resolution of $%
\pm 9.9 rd/s$.}
\label{PD-y3}
\end{figure}

For the three components of the plasma signal we have no way to
directly verify the accuracy of the tomographic estimates, because the
computed phase derivative is not exempt from noisy corruption. But in any
case the corresponding spectrogram plots show (Figs.\ref{PD-y1}, \ref%
{PD-y2-partial} and \ref{PD-y3}) that the tomogram allows for a good time
resolution and in no case departs from the approximate values obtained in
the corresponding spectrograms.

\begin{figure}[htb]
\begin{center}
\includegraphics[height=5cm,width=9cm]{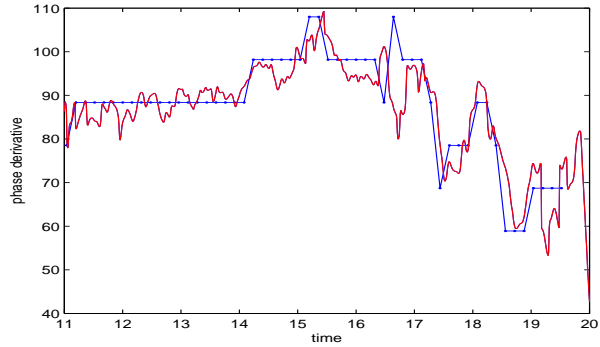}
\end{center}
\caption{For last part ($t\in \left[11,20\right]s$) of the second component
of the reflectometry signal, corresponding to the reflexion on the plasma : 
phase derivative obtained by  \textbf{TDM+LM} ($LM_{15}$). The
dots are the maximum of a moving window FFT spectrogram with a resolution of $\pm 9.9 rd/s$ .}
\label{PD-y2-partial}
\end{figure}

\section{ Remarks and conclusions}

The tomographic technique for component analysis and computation of the
phase derivative seems to provide an useful tool for the analysis of
reflectometry signals. The component separation technique contains more
information than the classical filtering techniques that have been used in
the past. In addition, the \textbf{TDM} method of phase derivative
calculation associated to \textbf{LM} filtering compares favorably with those
obtained by spectrograms. Nevertheless, a few issues must be addressed:

1. How many components should be separated? \cite{Briolle1}

From the tomogram itself one must decide how many components should be
extracted from the signal. From the analysis of a great number of reflectometry signals
it turned out that in some cases the third component, corresponding to the
multireflections, was very weak. Maybe, in this case, only two components
should be extracted.

2. Separation of the components: for which $\theta _{0}$ should the
separation be performed?

For $\theta \approx 0$, the spectrum $\left\{ c_{n}(\theta )\right\} $ is
very close to the time representation of the signal. Then, the coefficients $%
c_{n}(\theta )$ are almost all different from $0$, and it is not possible to
make the separation of the components. For $\theta \approx \frac{\pi }{2}$,
the spectrum $\left\{ c_{n}(\theta )\right\} $ is very close to the
frequency representation of the signal. Then, many coefficients $%
c_{n}(\theta )$ are equal to $0$ and the separation can be performed. But it
is not the best choice. The best choice for $\theta _{0}$ is where the
spectrum still has many non null coefficients $c_{n}(\theta )$ and where it
is still possible to make the separation by looking for concentrations of
the tomographic probability. In the case of the reflectometry signals, the
best choices seem to be around $\theta _{0}=\frac{3\pi }{10}$ (see Fig.\ref%
{Fi:reflec-cn-3pi_10-partial}). \newline

3. How is the phase derivative $\partial _{t}\phi (t)$ extracted?

3.1 First, one uses the time representation of the components to decide if
all parts of the signal are relevant, or if some of it is just filtered
noise (this is the case for the initial time interval in the second component of the reflectometry
signals)

3.2 Then, extract the phase derivative using \textbf{TDM} on the tomogram
for $\theta _{0}$ (For the first component of the reflectometry signal this
was sufficient to extract the phase derivative which is almost constant).

3.3 The use of \textbf{LM} filtering on the phase derivative can be relevant.
For the second and third component it seems necessary to apply a $LM_{15}$
low pass filter on the phase derivative.

4. The reflection on the porthole can be used to calibrate the measurements.
This reflection can be detected after the time $\tau _{1}$ corresponding to
the traveling wave from the emitting antenna to the receiver antenna. The
group delay $\tau _{g1}$ of the first reflection, computed from the phase
derivative should be equal to $\tau _{1}$.

The calibration of the measurements can be done by shifting the group delay $%
\tau _{g}$, obtained for each component from the phase derivative, by the
quantity $\Delta \tau =\tau _{1}-\tau _{g1}$.

\textbf{Acknowledgements:} The work reported in this paper is an ongoing
collaboration between the Center for Theoretical Physics (CNRS) - Marseille,
the Department of Research on Controlled Fusion (CEA) - Cadarache and the
Instituto de Plasmas e Fus\~{a}o Nuclear (IST) - Portugal. We acknowledge
financial support from Euratom/CEA (Contract No. V3517.001) and Euratom
Mobility. We are also grateful to F. Clairet, from Cadarache, for given us access 
to the reflectometry data.

\end{document}